\documentclass[epj]{svjour}

\usepackage{graphicx}
\usepackage{fancyhdr}
\usepackage{amsmath}
\def\makeheadbox{{
 }}
\newcommand{\DREG}{{\sc dreg}}
\newcommand{\DRED}{{\sc dred}}
\newcommand{\SUSY}{{\sc susy}}
\newcommand{\drbar}{\ensuremath{\overline{\sc DR}}}
\newcommand{\msbar}{\ensuremath{\overline{\sc MS}}}
\newcommand{\psib}{\overline{\psi}}
\newcommand{\epscalar}{$\varepsilon$-scalar}

\newcommand{\asDRbar}{\alpha_s^{\overline{\rm DR}}}
\newcommand{\asMSbar}{\alpha_s^{\overline{\rm MS}}}
\newcommand{\dd}{{\rm d}}
\newcommand{\reference}[1]{Ref.~\cite{#1}}
\newcommand{\ZsDRbar}{Z_s^{\overline{\rm DR}}}

\newcommand{\mDRbar}{m^{\overline{\rm DR}}}
\newcommand{\mMSbar}{m^{\overline{\rm MS}}}
\newcommand{\apiDR}{\frac{\asDRbar}{\pi}}

\def\alphadot{\dot\alpha}
\def\sigb{\overline{\sigma}}
\def\sigp{\sigma^{\prime}}

\def\mytitle{My title} 
\def\myauthors{My name}  
\def\mytype{My type of session}
\def\mysession{My session}

\def\mytitle{Dimensional Reduction Applied to Non-\SUSY{} Theories}
\def\myauthors{Philipp Kant}
\def\mytype{Contributed Talk}
\def\mysession{Theoretical Models}

\begin{document}
\title{Dimensional Reduction Applied to Non-\SUSY{} Theories}
\author{Philipp Kant
\thanks{\emph{Email:} kantp@particle.uni-karlsruhe.de}%
}                     
\institute{Institut f{\"u}r Theoretische Teilchenphysik, Universit\"at Karlsruhe}
%
\date{}
\abstract{
We consider regularisation of a Yang-Mills theory by
Dimensional Reduction (\DRED{}). In particular, the anomalous dimensions of
fermion masses and gauge coupling are computed to four-loop order.
We put special emphasis on the treatment of evanescent couplings which
appear when \DRED{} is applied to non-supersymmetric theories. We highlight
the importance of distinguishing between the evanescent and the real
couplings.
Considering the special case of a Super-Yang-Mills theory, we find that
Dimensional Reduction is sufficient to preserve Supersymmetry in our
calculations.
} 
\maketitle
%

\section{Introduction}

When calculating higher orders in perturbation theory, Dimensional
Regularisation (\DREG{})~\cite{'tHooft:1972fi,Bollini:1972ui} is the regularisation procedure of
choice.  Its convenience stems from the fact that it automatically preserves gauge
invariance: the finite part of the effective action satisfies the Ward
identities of the gauge symmetry, without the need to introduce
additional finite local counter\-terms.

When applied to supersymmetric theories, however, the Ward identities of
supersymmetry (\SUSY{}) are violated by the use of \DREG{}.  That is because
invariance of a given action under supersymmetry transformations only
hold for specific values of the space-time dimension, and \DREG{} alters
this value.

Dimensional Reduction (\DRED{})\cite{DRED} was proposed as a way of
reconciling \DREG{} and \SUSY{}.  The essence of this method is to
restrict the momenta to a $D$-dimensio\-nal subspace of the
$4$-dimensional space-time, while keeping all vector fields
$4$-dimensional.  Thus, the momentum integrals can be regularised
without meddling with the number of degrees of freedom of the gauge
fields, which is what breaks \SUSY{} in \DREG{}.

In the present talk, we will discuss the application of \DRED{} to a
Yang-Mills Theory with arbitrary gauge group. In particular, we will
outline the calculation of renormalisation group coefficients in a gauge
theory with fermions, using \DRED{} with minimal subtraction, which is
known as the \drbar{} scheme.  We will emphasise on the non-supersymmetric
case and the subleties that arise therein, namely the appearance of
evanescent couplings.  The calculations have been done up to the four-loop
order.

The correct application of \DRED{} to non-supersym\-metric theories is an
important issue in phenomenological studies when one wants to connect
parameters in a supersymmetric theorie valid at high energies with their
counterparts in a non-supersymmetric low-energy effective theory. A
recent example is the treatment of the running of the strong coupling
constant and the bottom mass in the {\sc mssm} in \reference{Harlander:2007wh}.

\section{Evanescent Couplings}

Consider a non-abelian gauge theory with gauge fields $W_{\mu}^a$ 
and a multiplet of  
two-component fermions $\psi_{\alpha}^{A}(x)$ transforming according to  a 
representation $R$ of the gauge group ${\cal G}$. 

The bare Lagrangian density, with covariant gauge fixing and ghost $(C,
C^*)$ terms is  
\begin{eqnarray}
	L_B &=& -\frac{1}{4}G^2_{\mu\nu} 
	- \frac{1}{2\alpha}(\partial^{\mu}W_{\mu})^2 
	+ C^{a*}\partial^{\mu}D_{\mu}^{ab}C^b \nonumber\\
	&&+ i\psib_{\alphadot A}\sigb^{\mu\alphadot\alpha}(D_{\mu})^{A}{}_B\psi^{B}_{\alpha} 
\end{eqnarray}
where 
\begin{equation}
G^a_{\mu\nu} = \partial_{\mu}W_{\nu}^a - \partial_{\nu}W_{\mu}^a 
+ gf^{abc}W_{\mu}^b W_{\nu}^c
\end{equation}
and 
\begin{equation}
(D_{\mu})^{A}{}_B = \delta^{A}{}_B\partial_{\mu} 
- ig (R^a )^{A}{}_BW_{\mu}^a .
\end{equation}
For the case when the theory admits a gauge invariant fermion mass 
term we will have $L_B \to L_B + L_B^m$, where 
\begin{equation}
L_B^m = \frac{1}{2} m_{AB}\psi^{\alpha A}\psi^B_{\alpha} 
+ \hbox{c.c.} 
\end{equation}

Applying \DRED{} amounts to imposing that all field variables
depend only on $D$ out of $4$ space-time dimensions, where  $D = 4 -
2\epsilon$. We can then make the decomposition 
\begin{equation}
W_{\mu}^a(x^j ) = W_i^a (x^j ) \oplus W_{\sigma}^a(x^j )
\end{equation}
where $\mu$ is an index of a $4$-dimensional, $i$ and $j$ are indices of
a $D$-dimensional, and $\sigma$ is an index of a $2\epsilon$-dimensional
vector space. An explicit construction of these vector spaces can be
found in \reference{Stockinger:2005gx}.
The Lagrange density then takes the form
\begin{equation}
L_B = L _B^d + L_B^{\epsilon} 
\end{equation}
where
\begin{equation}
L _B^d = -\frac{1}{4}G^2_{ij} -\frac{1}{2\alpha}(\partial^{i}W_{i})^2 +
C^{*}\partial^{i}D_{i}C + 
i\psib\sigb^i D_{i}\psi 
\label{eq:AD}
\end{equation}
and 
\begin{eqnarray}
 L_B^{\epsilon} &=& \frac{1}{2}(D_i W_{\sigma})^2 
- g\psib\sigb_{\sigma}R^a\psi W_{\sigma}^a \nonumber\\
&&-\frac{1}{4}g^2 f^{abc}f^{ade}W^b_{\sigma}W^c_{\sigp}W^d_{\sigma}W^e_{\sigp}.
\label{eq:AE}
\end{eqnarray}

Under gauge transformations, the $W_{\sigma}$-fields transform as
scalars, and they are commonly called \epscalar{}s.  Also, each term in
$L_B^{\epsilon}$ is separately invariant under gauge transformations,
so there is no reason to expect the form of Eq.~(\ref{eq:AE}) to be
preserved under renormalisation -- except in the case of a
supersymmetric theory, where invariance under supersymmetry
transformations requires $L_B^{\epsilon}$ to take the form of
Eq.~(\ref{eq:AE}). In general, the coupling of the 
\epscalar{}s to the fermions will not be governed by the gauge coupling
$g$, but by a different coupling $g_e$, called an evanescent coupling.

The case of the quartic \epscalar{} interaction is even more
complicated. Gauge invariance does not require the $f$-$f$ tensor
structure, but allows the quartic coupling to take the form 
\begin{equation}
-\frac{1}{4} \sum_{r=1}^p \lambda_r
            H^{abcd}_rW^a_{\sigma}W^c_{\sigp}W^b_{\sigma}W^d_{\sigp}
\end{equation}
where $H^{abcd}$ are tensors which  are non-vanishing when symmetrised
with respect to 
$(ab)$ and $(cd)$ interchange.  The number $p$ of such tensors which are
linearly independent depends on the group ${\cal G}$ and can be up to
$4$.  For instance, one could choose
\begin{eqnarray}
	2H_1 &=&  f^{ace}f^{bde} + f^{ade}f^{bce} \,,\nonumber\\
	2H_2 &=&  \delta^{ab}\delta^{cd}\,,\nonumber\\
	2H_3 &=&  \delta^{ac}\delta^{bd} +
	\delta^{ad}\delta^{bc} \,,\nonumber\\
	2H_4 &=&  f^{aef}f^{bfg}f^{cgh}f^{dhe} +
	f^{aef}f^{bfg}f^{dgh}f^{che}\,.
\end{eqnarray}

Corresponding to these evanescent couplings, we define the coupling constants 
\begin{equation}
\asDRbar{} = \frac{g^2}{4\pi}\,,\qquad
\alpha_e = \frac{g_e^2}{4\pi}\,,\qquad
\eta_r = \frac{\lambda_r}{4\pi}\,.
\label{eq:alpha}
\end{equation}

\section{Relating the \drbar{} to the \msbar{} scheme}

Considering two viable renormalisation schemes, it is possible to
translate calculations done in one scheme to the other scheme by finite
shifts of the renormalised parameters. 
In \reference{Harlander:2006rj}, we derived the relation between
$\asDRbar$ and $\asMSbar$ at 
the two-loop level using a method mentioned in \reference{Bern:2002zk}, which
relies on the fact that the value of $\alpha_s$ in a physical
renormalisation scheme should not depend on the regularisation procedure:
\begin{eqnarray}
  \alpha_s^{\rm ph} &=& \left(z_s^{\rm ph,X}\right)^2 \alpha_s^{\rm X}\,,\qquad
  z_s^{\rm ph,X} = Z_s^{\rm X}/Z_s^{\rm ph,X} \,,\nonumber\\
	&& \text{where } 
  {\rm X} \in \{\overline {\rm MS},\overline {\rm DR}\}\nonumber\\
  \Rightarrow
  \asDRbar &=& \left(\frac 
	   {Z_s^{\rm ph,\overline{\rm DR}}\,Z_s^{\overline{\rm MS}}}
	   {Z_s^{\rm ph,\overline{\rm MS}}\,Z_s^{\overline{\rm DR}}}
	   \right)^2
  \,\asMSbar\,,
\end{eqnarray}
where $Z_s^{{\overline {\rm MS}}/{\overline {\rm DR}}}$ are the charge
renormalisation constants using minimal
subtraction in \DREG{}/\DRED{}. For $Z_s^{{\rm ph},\overline{\rm
		MS}/\overline{\rm DR}}$, on the other hand, we used \DREG{}/\DRED{}
combined with a physical renormalisation condition. 

The two-loop result of \reference{Harlander:2006rj} reads, for the case of QCD, 
\begin{equation}
  \asMSbar{} = \asDRbar{}\left[1-\frac{\asDRbar{}}{4\pi}
    - \frac{5}{4}\left(\frac{\asDRbar{}}{\pi}\right)^2
    + \frac{\asDRbar{}\alpha_e}{12\pi^2}\,n_f\right]\,.
  \label{eq::asMS2DR}
\end{equation}
At the three-loop level, the quartic \epscalar{} interaction starts to
contribute. In \reference{Harlander:2006xq}, the three-loop term in the
conversion relation was calculated for the case of QCD, and in \reference{Jack:2007ni}
it was possible to calculate it for an arbitrary gauge group.

In addition to the conversion formulae for $\alpha_s$, also conversion
formulae for the quark mass in the \msbar{} and \drbar{} scheme have been found in
\reference{Harlander:2006rj,Harlander:2006xq,Jack:2007ni}, using the same technique.

\section{Renormalisation Group Coefficients}

The dependence of the coupling constants (\ref{eq:alpha}) and the quark
mass on the renormalisation scale $\mu$ is given by their
$\beta$ functions
\begin{eqnarray}
	\beta_s^{\drbar}(\asDRbar,\alpha_e,\{\eta_r\}) &=& \mu^2\frac{\dd}{\dd\mu^2}\asDRbar\,,\nonumber\\
	\beta_e(\asDRbar,\alpha_e,\{\eta_r\}) &=& \mu^2\frac{\dd}{\dd\mu^2}\alpha_e\,,\nonumber\\
	\beta_{\eta_r}(\asDRbar,\alpha_e,\{\eta_r\}) &=& \mu^2\frac{\dd}{\dd\mu^2}\eta_r\,,\nonumber\\
	\gamma_m^{\drbar}(\asDRbar,\alpha_e,\{\eta_r\}) &=& \mu^2\frac{\dd}{\dd\mu^2}m\,,
\end{eqnarray}
which can be calculated if one knows the corresponding renormalisation
constants. For instance, 
\begin{eqnarray}
  \lefteqn{\beta_s^{\drbar}(\asDRbar,\alpha_e,\{\eta_r\}) = }
 \nonumber\\
   - &&\left(\epsilon \apiDR + 2 \frac{\asDRbar}{\ZsDRbar} 
    \frac{\partial \ZsDRbar}{\partial \alpha_e} \beta_e
    + 2 \frac{\asDRbar}{\ZsDRbar}\sum_r 
    \frac{\partial \ZsDRbar}{ \partial \eta_r} \beta_{\eta_r} 
  \right)\nonumber\\
  &&\quad\left(1+  2 \frac{\asDRbar}{\ZsDRbar} \frac{\partial \ZsDRbar}
    {\partial \asDRbar}\right)^{-1}    
  \,.
\end{eqnarray}

However, Eq.~(\ref{eq::asMS2DR}) and its extension to three loop-level
allows to transfer the gauge and fermion mass $\beta$ functions from the
\msbar{} scheme, where they are known to the four-loop level
\cite{vanRitbergen:1997va,Chetyrkin:1997dh,Vermaseren:1997fq,Czakon:2004bu}
to the \drbar{} scheme via
\begin{eqnarray}
  \beta_s^{\drbar}
  &=& \beta_s^{\msbar}
  \frac{\partial \asDRbar}{\partial \asMSbar} +
  \beta_e \frac{\partial \asDRbar}{\partial \alpha_e} +
  \sum_r \beta_{\eta_r} \frac{\partial \asDRbar}{\partial \eta_r}
  \,,
  \nonumber\\
  \gamma_m^{\drbar} &=&
  \gamma_m^{\msbar} \frac{\partial \ln \mDRbar}{\partial \ln \mMSbar}  
  + \frac{\pi \beta_s^{\msbar}}{\mDRbar}
  \frac{\partial \mDRbar}{\partial \asMSbar}
  \nonumber\\&&
  + \frac{\pi \beta_e}{\mDRbar}
  \frac{\partial \mDRbar}{\partial \alpha_e}
  + \sum_r \frac{\pi \beta_{\eta_r}}{\mDRbar}  
  \frac{\partial \mDRbar}{\partial \eta_r}
  \,.
  \label{eq::DRED-DREG}
\end{eqnarray}
In \reference{Jack:2007ni}, a four-loop result for $\beta_s^{\drbar}$ and
$\gamma_m^{\drbar}$ was derived using Eq.~(\ref{eq::DRED-DREG}). In
addition to the \msbar{} result, the following building blocks were
needed: since the dependence of $\asDRbar$ and
$\mDRbar$ on $\alpha_e$ starts at two- and one-loop
order~\cite{Harlander:2006rj}, respectively, $\beta_e$ is needed
up to the three-loop level. On the other
hand, both $\asDRbar$ and $\mDRbar$ depend on $\eta_r$ starting from
three loops and consequently only the one-loop term of $\beta_{\eta_r}$
enters in Eq.~(\ref{eq::DRED-DREG}).

It should be noted that the $\beta$ functions of the evanescent
couplings differ from the gauge $\beta$ function starting already at the
one-loop level. Hence, it is not possible to identify the evanescent
couplings with the gauge coupling.

\section{Super-Yang-Mills Checks}

Starting with a Yang-Mills theory, it is possible to construct a
(supersymmetric) Super-Yang-Mills theory by putting the fermions in the
adjoint representation. This is useful for applying checks to the
results of \reference{Harlander:2006rj,Harlander:2006xq,Jack:2007ni}: in a
supersymmetric theory, the evanescent 
coupling $g_e$ must equal the gauge coupling, so their $\beta$ functions
should also be the same. We checked this equality to the three-loop
level, which is a strong check on our calculation and also invalidates
an earlier claim \cite{Avdeev:1982np} that $\beta_s$ and $\beta_e$ would
differ at the three-loop level in a Super-Yang-Mills theory. In
\reference{Avdeev:1982np}, the inequality of the $\beta$ functions was
interpreted as an example of \SUSY{} breaking by \DRED{}. 

In \reference{Jack:1998uj}, the four-loop gauge $\beta$ function of a
Super-Yang-Mills theory was presented. This provided another check to
our calculations, and we did find agreement.

\section{Discussion}

We demonstrated that Dimensional Reduction is a viable regularisation
procedure even in the non-supersym\-metric case and derived explicit
conversion formulae for the gauge coupling and quark mass between the
\msbar{} and the \drbar{} scheme. We explained the appearance of
evanescent couplings and emphasised that they cannot be identified with
the gauge coupling, since the corresponding $\beta$ functions differ.

We calculated gauge and fermion mass $\beta$ functions for arbitrary
gauge theories and applied various checks in the special case of
supersymmetric theories.

\end{document}